\documentstyle[preprint,aps]{revtex}
\preprint{NSF-ITP-95-125}
\draft
\title{Localization Effects in ac-driven Tight-Binding Lattices}
\author{Martin Holthaus~\footnote{Electronic address:
    holthaus@stat.physik.uni-marburg.de}}
\address{Fachbereich Physik der Philipps-Universit\"at,	\\
    Renthof 6, D--35032 Marburg, Germany}
\author{Daniel W.~Hone~\footnote{Electronic address:
    hone@itp.ucsb.edu}}
\address{Institute for Theoretical Physics and QUEST, \\
    University of California at Santa Barbara, Santa Barbara, CA 93106}
\date{\today}
\narrowtext
\begin{document}
\maketitle
\begin{abstract}
We study coherent dynamics of tight-binding systems interacting with static
and oscillating external fields. We consider Bloch oscillations and
Wannier-Stark localization caused by dc fields, and compare these effects
to dynamic localization that occurs in the presence of additional ac fields.
Our analysis relies on quasienergy eigenstates, which take over the role
of the usual Bloch waves. The widths of the quasienergy bands depend
non-monotonically on the field parameters. If there is lattice disorder,
the degree of the resulting Anderson localization is determined by the ratio
of disorder strength and quasienergy band width. Therefore, the localization
lengths can be controlled, within wide ranges, by adjusting the ac amplitude.  
Experimental realizations of our model systems are given by semiconductor
superlattices in far-infrared laser fields, or by ultracold atoms in
modulated standing light waves.  In both cases the system parameters, as
well as the field amplitudes and frequencies, are readily accessible to
experimental control, suggesting these as highly attractive candidates for
systematic study of localization phenomena.
\end{abstract}
%\preprint{}

\section{Introduction: The ``Renormalization'' of the Land\'e $g$-Factor}

More than 25 years ago Haroche, Cohen-Tannoudji, Audoin, and Schermann
demonstrated experimentally that the Land\'{e} factor $g_F$ of atomic
hyperfine levels $F$ can be modified by an oscillating magnetic
field~\cite{HarocheEtAl70}. Working with Hydrogen and Rubidium atoms,
respectively, and applying an oscillating field $B_{1}\cos(\omega t)$
perpendicular to a static field $B$, the authors showed that the effective
value $g^{(eff)}_F$ of the Land\'e factor in the presence of both fields
is given by $g^{(eff)}_F = g_F J_0(g_F \mu_B B_1/\omega)$, where $\mu_B$ is
the Bohr magneton, and $J_0$ is the ordinary Bessel function of order zero.
Similar results have also been published by Chapman~\cite{Chapman70}
and Yabuzaki et al.~\cite{YabuzakiEtAl72,YabuzakiEtAl74}.

This renormalization of the Land\'e factor by oscillating fields is closely
related to the subject of the present paper: the renormalization of the band
structure in tight-binding lattices caused by external ac electric fields,
and the resulting possibility of manipulating tunneling and localization effects
by suitably tuning the ac amplitudes. In order to elucidate how an oscillating
field can affect the Land\'e factor, and to prepare for the following discussion
of tight-binding systems, we consider an electron in a state with total
angular momentum $F = 1$. We assume that a static magnetic field of strength
$B$ is applied in the $z$-direction, together with an additional static
field $B_0$ and an oscillating field $B_1\cos(\omega t)$ in the $x$-direction.
The Hamiltonian, restricted to the $F=1$ subspace, can then be written as
\begin{equation}
H(t) = g_1\mu_B B \left( \begin{tabular}{ccc}
	    1 & 0 & 0  \\
	    0  & 0 & 0  \\
	    0  & 0 & -1 \\ \end{tabular} \right)
\; + \; g_1\mu_B\left[B_0 + B_1\cos(\omega t)\right] \, \frac{1}{\sqrt{2}}
\left(\begin{tabular}{ccc}
            0  & 1 & 0  \\
	    1  & 0 & 1  \\
	    0  & 1 & 0  \\ \end{tabular} \right)	\; .		
\label{HEX}
\end{equation}

Because $H(t)$ is periodic in time with period $T = 2\pi/\omega$, there
are Floquet states~\cite{Shirley65,Zeldovich67,Ritus67}
\begin{equation}
\psi_{\alpha}(t) = u_{\alpha}(t)\exp(-i\varepsilon_{\alpha} t) \; , 
\end{equation}
where the $T$-periodic Floquet functions $u_{\alpha}(t)$ solve the eigenvalue
equation
\begin{equation}
\left( H(t) - i\partial_t \right) u_{\alpha}(t) =
    \varepsilon_{\alpha} u_{\alpha}(t)	\; .	    \label{EIG}
\end{equation}
The eigenvalues $\varepsilon_{\alpha}$ are called quasienergies.

Now there is a simple, but important detail to note. Suppose that $u_j(t)$
is a $T$-periodic solution to~(\ref{EIG}), with quasienergy $\varepsilon_{j}$.
Then, for every positive or negative integer $m$, the function
$u_j(t)\exp(im\omega t)$ is also a $T$-periodic eigensolution, with
quasienergy $\varepsilon_j + m\omega$. 
Thus, the quasi\-energy spectrum consists of ``Brillouin zones'' of width
$\omega$, and the index $\alpha$ in~(\ref{EIG}) should be regarded as a
double-index: $\alpha = (j,m)$. The eigenvalue problem~(\ref{EIG}) is defined
in an extended Hilbert space of $T$-periodic functions~\cite{Sambe73},
with scalar product
\begin{equation}
\langle\langle \; \cdot \; | \; \cdot \; \rangle\rangle \; = \;
\frac{1}{T}\int_0^{T} \! {\mbox d}t \,
\langle \; \cdot \; | \; \cdot \; \rangle \; .
\label{SCA}
\end{equation}
This is the usual scalar product combined with time-averaging. {\em All}
eigenfunctions $u_j(t)\exp(im\omega t)$ are needed for the completeness
relation in the extended Hilbert space.  

On the other hand, all solutions that differ only by a factor
$\exp(im\omega t)$ fall into the same class, in that they represent the
same physical state $\psi_j(t) = u_j(t)\exp(-i\varepsilon_j t)$.
An arbitrary solution $\psi(t)$ to a time-dependent Schr\"odinger equation
with a $T$-periodic Hamiltonian $H(t)$ can be expanded according to
\begin{equation}
\psi(t) = \sum_j a_j u_j(t)\exp(-i\varepsilon_j t)  \; ,
\end{equation}
where $j$ labels eigensolutions of~(\ref{EIG}) that correspond to different
states.
This expansion shows that for quantum systems governed by a time-periodic
Hamiltonian the Floquet states and quasienergies play the same role as energy
eigenstates and energies do in the static case.
The coefficients $a_j$ do {\em not} depend on time. Note that the Fourier
index $m$ does not appear: only one representative $u_j(t)$ of each class of
eigenfunctions is required here. For this reason, we will not use the
double-index notation in the following, and merely indicate the
Brillouin-zone structure of the quasienergy spectrum by adding
``$\bmod \, \omega$'' to the corresponding formulae.
 
We now compute the quasienergies for the Hamiltonian~(\ref{HEX})
perturbatively. A unitary transformation yields 
\begin{equation}
\widetilde{H}(t) = g_1\mu_{B} B \, \frac{1}{\sqrt{2}}\left(
\begin{tabular}{ccc}
	0          & 1	          & 0	         \\
	1          & 0            & 1		 \\
	0          & 1            & 0            \\ \end{tabular} \right)
\; + \; g_1\mu_B\left[B_0 + B_1\cos(\omega t)\right] \left(\begin{tabular}{ccc}
        1          & 0	          & 0		 \\
	0	   & 0	          & 0		 \\
	0          & 0	          & -1           \\ \end{tabular} \right) \; .
\end{equation}
We will denote the first term on the right hand side as $H^{(B)}$.
If $B=0$, the wave functions
\[
\psi_1(t) = \left(\begin{tabular}{c} 1 \\ 0 \\ 0 \\ \end{tabular} \right)
    \exp\!\left(-ig_1\mu_B(B_0 t + \frac{B_1\sin(\omega t)}{\omega})\right)
\;\;\; , \;\;\;
\psi_2(t) = \left(\begin{tabular}{c} 0 \\ 1 \\ 0 \\ \end{tabular} \right)
\;\;\; , 
\]
\begin{equation}
\psi_3(t) = \left(\begin{tabular}{c} 0 \\ 0 \\ 1 \\ \end{tabular} \right)
    \exp\!\left(+ig_1\mu_B(B_0 t + \frac{B_1\sin(\omega t)}{\omega})\right)
\label{WAV}
\end{equation}
solve the time-dependent Schr\"odinger equation. Obviously, these are
Floquet states with quasienergies $\varepsilon_M = g_1 \mu_B B_0 \cdot M$
for $M = -1, 0, +1$, to be taken modulo~$\omega$.

If $B$ is small, $B \ll B_0 \approx B_1$, the matrix $H^{(B)}$
can be considered as a perturbation. Perturbation theory for quasienergy
eigenstates in the extended Hilbert space works exactly like the well-known
Rayleigh-Schr\"odinger perturbation theory for energy eigenstates, provided
the proper scalar product~(\ref{SCA}) is employed. In our case,
the ``unperturbed'' Floquet functions $u_j(t)$ can be identified as
the $T$-periodic parts of the wave functions~(\ref{WAV}).
If there is no integer $n$ such that $g_1\mu_B B_0 = n\omega$, then there
is no resonance, and we can resort to nondegenerate perturbation theory.
For small $B$ it suffices to calculate the diagonal elements of the
perturbation. Since
\begin{equation}
\langle\langle u_j(t) | H^{(B)} | u_j(t) \rangle\rangle = 0 \qquad \qquad
(j = 1,2,3) \;\ , 
\end{equation}
there is no first-order correction to the quasienergies.

The situation is different if $g_1\mu_B B_0 = n\omega$ for some integer $n$
--- i.e., if the energy of $n$ photons\footnote{Since the ac field is treated
as a classical external force, rather than a quantized field, the term
``photon'' is, strictly speaking, not justified here. We use it nonetheless
to express the resonance condition in a form that can be easily remembered.}
coincides with the energy difference between adjacent Zeeman states.
Then the Brillouin-zone structure of the quasienergy spectrum becomes crucial:
the Floquet eigenfunctions
\[
u_1(t) = \left(\begin{tabular}{c} 1 \\ 0 \\ 0 \\ \end{tabular} \right)
    \exp\!\left(-ig_1\mu_B\frac{B_1\sin(\omega t)}{\omega}\right)
\;\;\; , \;\;\;
u_2(t) = \left(\begin{tabular}{c} 0 \\ 1 \\ 0 \\ \end{tabular} \right)
    e^{in\omega t}
\;\;\; , 
\]
\begin{equation}
u_3(t) = \left(\begin{tabular}{c} 0 \\ 0 \\ 1 \\ \end{tabular} \right)
    \exp\!\left(+ig_1\mu_B\frac{B_1\sin(\omega t)}{\omega}\right)
    e^{2in\omega t}
\end{equation}
are degenerate. Therefore one has to diagonalize the matrix of the perturbation
$H^{(B)}$ within the degenerate subspace. Using the generating function
for the ordinary Bessel functions $J_k$ of integer order,
\begin{equation}
e^{iz\sin(\varphi)} = \sum_{k=-\infty}^{+\infty} e^{ik\varphi} J_k(z) \; ,
\label{GEN}
\end{equation}
this matrix becomes (with $j,k = 1,2,3$)
\begin{equation}
\langle\langle u_j(t) | H^{(B)} | u_k(t) \rangle\rangle =
    \frac{(-1)^n}{\sqrt{2}} g_1 \mu_B B \,
    J_{n}\!\left(\frac{g_1\mu_B B_1}{\omega}\right) \delta_{j,k\pm1}	\; .
\end{equation}
Hence, the quasienergies are given by
\begin{equation}
\varepsilon_{M} = (-1)^n J_{n}\!\left(\frac{g_1\mu_B B_1}{\omega}\right)
    g_1\mu_B B \cdot M \quad \bmod \omega \; ,
\end{equation}
for $M = 0,\pm 1$. Thus, in effect the $g$-factor is ``renormalized'' by the
ac field:
\begin{equation}
g_1 \; \longrightarrow \; (-1)^n J_{n}\!\left(
    \frac{g_1 \mu_B B_1}{\omega}\right) \cdot g_1 \; \equiv \; g_1^{(eff)} \; .
\label{REG}
\end{equation}

Originally, the renormalization of the atomic $g$-factors had been explained
within the dressed-atom picture~\cite{CohenTannoudji68,CohenTannoudji92},
which amounts to treating both the atom and the oscillating field quantum
mechanically. Within the Floquet picture the field is not quantized, but is
regarded as an external force acting on the atom. The two approaches give
identical results, as long as the ac fields are sufficiently
intense~\cite{Shirley65}. The line of reasoning applied to the above example
is essentially a formalized version of a semiclassical argument given earlier
by Pegg and Series~\cite{PeggSeries70}.  

The lesson to learn from this example is that resonant oscillating
fields can substantially alter the level structure of atoms or molecules.
Similarly, ac fields can modify the band structure in periodic potentials.
This effect is of possible relevance for present experiments on
semiconductor superlattices exposed to far-infrared
radiation~\cite{FEL1,IgnatovEtAl94,FEL2,FEL3}:
ideally, one can alter the quasienergy (mini-)bandwidths in these systems
by tuning the amplitude of the radiation field, and thereby control tunneling
and localization phenomena in the laboratory. In particular, experiments of
this kind might shed new light on Anderson localization.   

In the following, we explain this possibility in more detail.
We first discuss a simple two-level problem, which already exhibits many of
the central features, and then move on to one-dimensional
tight-binding lattices. We review the phenomena of Bloch oscillations and
Wannier-Stark localization occuring in the presence of dc electric fields,
and the dynamic localization that one finds in combined ac and dc fields. 
A crucial step then follows: we introduce disorder into the tight-binding
model, and demonstrate that the parameter which determines the degree of
disorder-induced localization is the ratio of disorder strength and the
{\em quasi\/}energy bandwidth. The case of an isolated defect is treated
analytically; numerical results are given for randomly disordered lattices.
The role of interband effects is discussed within the scope of a two-band
model. The paper then closes with a brief discussion and outlook.

\section{The Two-Level System}

Let us first consider a rudimentary ``lattice'' with just two sites,
each of which supports only a single quantum state --- i.e., a two-level
system, which interacts with an external dc electric field of strength
$F_0$ and an additional ac field of strength $F_1$ and frequency $\omega$:
\begin{equation}
H_{tls}(t) = \frac{\Delta}{2}\sigma_{z}
+ \mu[F_0 + F_1\cos(\omega t)]\sigma_{x}  \; .
\label{TLS}
\end{equation}
The energy difference between the unperturbed states is denoted by
$\Delta$, $\mu$ is the dipole matrix element, and $\sigma_x$, $\sigma_z$
are the usual Pauli matrices. This Hamiltonian describes, for example,
an electron in a semiconductor double-well heterostructure under dc bias
in the presence of far-infrared
radiation~\cite{DakhnovskiiBavli93,GorbatsevichEtAl95},
provided the dynamics remain restricted to the lowest doublet of
energy eigenstates.

Evidently, $H_{tls}(t)$ is of the same type as the Hamiltonian~(\ref{HEX})
(but with two levels, or effective spin $1/2$ rather than spin $1$),
and the perturbative calculation of the quasienergies proceeds exactly along
the lines of the previous example. For $\Delta = 0$ the Floquet functions
take the form
\begin{equation}
u_{\pm}(t) = \frac{1}{\sqrt{2}} \left(\begin{array}{rr} 1 \\ \pm 1 \end{array}
\right)\exp[\mp i\frac{\mu F_{1}}{\omega}\sin(\omega t)] \;\;\; ,
\end{equation}
and the quasienergies are     
\begin{equation}
\varepsilon_{\pm} = \pm \mu F_0 \qquad \bmod \omega \;\;\; .
\end{equation}
If the ac field is not resonant, i.e., if $2\mu F_0 \ne n\omega$
for all integer $n$, these quasienergies remain unchanged to first
order in $\Delta/\omega$. However, if $2\mu F_0 = n\omega$ for some integer
$n$, then degenerate perturbation theory in the extended Hilbert space
yields
\begin{equation}
\varepsilon_{\pm} = \frac{n\omega}{2} \pm (-1)^n\frac{\Delta}{2}
    J_n\!\left(\frac{2\mu F_1}{\omega}\right) \; \qquad \bmod \omega \; .
\label{SHI} 
\end{equation}
If there is no dc field, $F_0 = 0$, the resonance condition is fulfilled
automatically with $n = 0$. In this case the two quasienergies cross when
$2\mu F_1/\omega$ is approximately equal to a zero of the Bessel function
$J_0$. This fact, which had already been derived by Shirley in
1965~\cite{Shirley65} and thoroughly studied
experimentally~\cite{YabuzakiEtAl74}, has sparked renewed interest recently,
after it was pointed out by
Grossmann et al.~\cite{GrossmannEtAl91a,GrossmannEtAl91b,Haenggi95}
that tunneling in one-dimensional symmetric double-well potentials can be
coherently suppressed by an oscillating force. Assuming that the wave packet
dynamics can be captured by a two-level approximation, an obvious necessary
condition for the suppression of tunneling is the crossing of the two
quasienergies emerging from the lowest doublet of energy eigenstates of the
unperturbed double well, and Shirley's formula~(\ref{SHI}) gives a reasonable
approximation to the parameters where the quasienergy crossing
occurs~\cite{LlorentePlata92,GrossmannHaenggi92}. 
The two-level approximation becomes insufficient when the ac frequency or
amplitude is so high that the field not only mixes the lowest two double well
eigenstates but also couples these states to higher doublets. Then the
quasienergy ``ground state'' splitting can become even {\em larger} than the
bare tunnel splitting $\Delta$. This fact enables fast, efficient, and
externally controllable population transfer in ac-driven double
wells~\cite{Holthaus92a}.
It should be pointed out, however, that the full quasienergy spectral
problem for periodically driven anharmonic oscillators is technically far
more demanding than the computation of spectra for $N$-level systems.
At present, it is even unknown whether a periodically forced oscillator
with a quartic anharmonicity has a singular or an absolutely
continuous quasienergy spectrum, if the amplitude of the driving force
is large~\cite{Howland92}.

\section{The Infinite Tight-Binding Lattice in dc Fields:
	Wannier-Stark Localization and Bloch Oscillations}

What happens to the approximate spectrum~(\ref{SHI}) when the number of
lattice sites is increased from two to infinity? This question leads us
to consider the tight-binding Hamiltonian 
\begin{equation}
H_{0} = -\frac{\Delta}{4}\sum_{\ell}\left( |\ell+1 \rangle \langle \ell|
    \, + \, |\ell \rangle \langle \ell+1| \right)
\label{TBS}
\end{equation}
with scalar potential interaction
\begin{equation}
H_{int}(t) = e [F_0 + F_1\cos(\omega t)] d \,
\sum_{\ell} |\ell\rangle \, \ell \, \langle \ell |  \;\; .
\label{HIN}
\end{equation}
The system defined by the Hamiltonian $H_0 + H_{int}(t)$
can be taken as a model for an electron in an ideal superlattice in
the presence of both terahertz radiation, linearly polarized along the growth
direction of the lattice, and additional dc bias. An ``atomic'' (Wannier)
state at the $\ell$-th site is denoted by $|\ell\rangle$, and $d$ is the
lattice constant. If there is no external electric field, $F_0 = F_1 = 0$,
then the Bloch waves
\begin{equation}
\chi_k = \sum_{\ell}e^{-ik\ell d} |\ell\rangle	\label{BLO}
\end{equation}
solve the stationary Schr\"odinger equation, $H_0\chi_k = E(k)\chi_k$,
with energy eigenvalues
\begin{equation}
E(k) = -\frac{\Delta}{2}\cos(kd) \; .    \label{EOK}
\end{equation}
Thus, $\Delta$ is the width of the resulting energy band.

If there is a dc field $F_0$, but no ac field, the Wannier-Stark states
\begin{equation}
\varphi_m = \sum_{\ell} J_{\ell-m}\!\left(\frac{\Delta}{2eF_0d}\right)
|\ell\rangle
\label{WAN}
\end{equation}
constitute a complete set of energy eigenstates~\cite{FukuyamaEtAl73}.
Their eigenvalues,
\begin{equation}
E_m = m \cdot e F_0 d  \; ,
\label{WSS}
\end{equation}
form the so-called Wannier-Stark ladder.
A particular Wannier-Stark state $\varphi_{m}$ is localized around the
$m$-th site, with a localization length of the order of $\Delta/(eF_0 d)$
lattice periods. The addition theorem for the Bessel functions $J_{k}(z)$
of integer order,
\begin{equation}
\sum_{k=-\infty}^{+\infty} J_{k}(z) J_{k+p}(z) e^{ik\alpha}
= J_{p}(2z\sin(\alpha/2))e^{ip(\pi-\alpha)/2}	\; ,
\label{ADI}
\end{equation}
directly reflects, for $\alpha = 0$, completeness and orthogonality of these
states. 
     
The propagator for the dc-biased lattice reads~\cite{FukuyamaEtAl73}
\begin{eqnarray}
K_{\ell,\ell'}(t) & = & \; \sum_m \langle \ell | \varphi_m \rangle
\langle \varphi_m | \ell'\rangle \exp(-iE_m t)
\nonumber \\ & = &
\sum_{m} J_{\ell-m}\!\left(\frac{\Delta}{2eF_0d}\right)
J_{\ell'-m}\!\left(\frac{\Delta}{2eF_0d}\right)\exp(-imeF_0dt)
\nonumber \\ & = &
J_{\ell-\ell'}\!\left(\frac{\Delta}{eF_0d}\sin\left(\frac{eF_0dt}{2}\right)
\right)\exp\!\left(i(\ell-\ell')\frac{\pi - eF_0dt}{2} -i\ell'eF_0dt\right) \; ,
\label{PRO}
\end{eqnarray}
where~(\ref{ADI}) has been used. Because the Wannier-Stark spectrum~(\ref{WSS})
is a ladder with equidistant level spacing $\Delta E = eF_0d$, the propagator
$K_{\ell,\ell'}(t)$ is periodic in time with the Bloch period
$T_{Bloch} = 2\pi/(eF_0d)$. Hence the motion of any wave packet is also
periodic with that same period. But there are important features of the
general dynamics beyond this simple periodicity.  To understand these it is
instructive to study the motion of an initially localized Gaussian wave packet:
\begin{equation}
\psi(t=0) \; = \; \sum_{\ell} f_{\ell}(0) |\ell\rangle
\end{equation}
with
\begin{equation}
f_{\ell}(0) = \left(\frac{d^2}{2\pi\sigma^2}\right)^{1/4}
\exp\left(-\frac{\ell^2d^2}{4\sigma^2} + ik_0\ell d\right) \; .
\label{GAU}
\end{equation}
Fourier-transforming the time-evolution equation
$f_{\ell}(t) = \sum_{\ell'} K_{\ell,\ell'}(t) f_{\ell'}(0)$,
and utilizing~(\ref{GEN}), one obtains
\begin{eqnarray}
\widehat{f}(k,t)
    & = & \sum_{\ell,\ell'}e^{-ik\ell d}K_{\ell,\ell'}(t) f_{\ell'}(0)
\nonumber
\\  & = & \exp\left[i\frac{\Delta}{eF_0d}\sin\left(\frac{eF_0dt}{2}\right)
    \cos\left(kd + \frac{eF_0dt}{2}\right)\right]
    \sum_{\ell'}e^{-i(k+eF_0t)\ell'd} f_{\ell'}(0)  \; .
\label{SUM}
\end{eqnarray}
This equation is still exact. If we now assume that $\sigma/d \gg 1$,
so that the spatial localization of the initial wave packet~(\ref{GAU})
is weak, then the summation over $\ell'$ can approximately be replaced by
integration, and we arrive at
\begin{equation}
\widehat{f}(k,t) \approx \left(\frac{8\pi\sigma^2}{d^2}\right)^{1/4}
    \exp\left[i\frac{\Delta}{eF_0d}\sin\left(\frac{eF_0dt}{2}\right)
    \cos\left(kd + \frac{eF_0dt}{2}\right)\right]
    e^{-\sigma^2(k - k_0 + eF_0t)^2}	\; .
\label{INT}
\end{equation}
At this point we can make contact with two familiar notions. First, it can be
seen that the center of the wave packet in $k$-space is given by
$\widehat{k}(t) = k_{0} - eF_0t$. Hence,
\begin{equation}
\frac{d}{dt} \widehat{k}(t) = -e F_0    \; .
\label{ACT}
\end{equation}
That is a special case of Bloch's famous ``acceleration theorem'': 
the rate of change of the wave packet's center in $k$-space is determined
by the external force~\cite{Bloch28}. Second, it should be realized that the
explicit $T_{Bloch}$-periodicity of the discrete sum~(\ref{SUM}) is apparently
lost if one replaces summation by integration. But the wave vector $k$ is
defined, because of spatial periodicity, only modulo $2\pi/d$, so this is also
true of the argument $(k - k_0 + eF_0t)$ appearing in the exponential
of~(\ref{INT}), restoring the temporal periodicity with the Bloch period.
The corresponding physics is that the wave packet is Bragg-reflected at the
edges of the Brillouin zone.   

Now we are interested in the evolution of the wave packet in real space:
\begin{equation}
f_{\ell}(t) = \frac{d}{2\pi} \int_{-\pi/d}^{+\pi/d} \! {\mbox d}k \,
    \widehat{f}(k,t)e^{ik\ell d}    \; .
\label{FTB}
\end{equation}
Since we have already stipulated that $\sigma/d \gg 1$, the momentum
distribution $\widehat{f}(k,t)$ is sharply localized around $\widehat{k}(t)$.
Therefore, it makes sense to expand the factor $\cos(kd + eF_0dt/2)$
in the exponential of~(\ref{INT}) around $\widehat{k}(t)$.
>From a linear expansion one finds
\begin{equation}
f_{\ell}(t) \approx \left(\frac{d^2}{2\pi\sigma^2}\right)^{1/4}
\exp\left(i(k_0 - eF_0t)\ell d - i\Phi(t)
-\frac{d^2}{4\sigma^{2}}(\ell - \widehat{\ell}(t))^2\right)
\end{equation}
with
\begin{equation}
\Phi(t) = \frac{\Delta}{2eF_0d}\left[\sin(k_0d - eF_0dt) - \sin(k_0d)\right]
\end{equation}
and
\begin{equation}
\widehat{\ell}(t) = \frac{\Delta}{2eF_0d}\left[\cos(k_0d - eF_0dt)
    - \cos(k_0d)\right]	\; .
\end{equation}
Thus, the center of the wave packet, $\widehat{x}(t) = \widehat{\ell}(t)d$,
oscillates in space with an amplitude that is inversely proportional to the
dc field strength $F_0$, and with temporal period $T_{Bloch}$. This is the
famous ``Bloch oscillation'', that was first discussed by
Zener~\cite{Zener34}.
 
A simple ``semiclassical'' explanation of these wave packet oscillations can
be given as follows. First one introduces a continuous position variable $x$
instead of the discrete lattice points $x_{\ell} = \ell d$. From the
``acceleration theorem''~(\ref{ACT}) one knows $\widehat{k}(t)$. Furthermore,
the group velocity $\widehat{v}(t)$ of the wave packet is determined
by the dispersion relation~(\ref{EOK}) pertaining to the unperturbed lattice:
\begin{equation}
\widehat{v}(t) = \left.\frac{{\mbox d}E(k)}{{\mbox d}k}\right|_{\widehat{k}(t)}
    = \frac{\Delta d}{2}\sin(\widehat{k}(t)d)   \; .
\end{equation}
Assuming $\widehat{x}(0) = 0$ and integrating, one finds for the position
$\widehat{x}(t)$ of the wave packet's center
\begin{equation}
\widehat{x}(t)
    = \frac{\Delta}{2eF_0}\left[\cos(k_0d - eF_0dt) - \cos(k_0d)\right] \; ,
\end{equation}
which agrees exactly with $\widehat{\ell}(t)d$ as derived above. The previous
quantum mechanical calculation shows that this semiclassical reasoning is
essentially correct as long as the spatial localization of the initial wave
packet is weak, $\sigma/d \gg 1$, so that the momentum distribution
$\widehat{f}(k,t)$ is sharply localized within a single Brillouin zone of width
$2\pi/d$. This assumption has not only been used when replacing sums by
integrals, but also in the expansion employed to evaluate~(\ref{FTB}). It
should be emphasized that the expansion around $\widehat{k}(t)$ is not an
expansion around a stationary point. It can, therefore, be justified only if
the Gaussian in~(\ref{INT}) is sharply peaked around $\widehat{k}(t)$. That
is the reason why we did not employ a second-order expansion, which would have
resulted in a periodic oscillation of the wave packet width.
But if the Gaussian is only moderately peaked, so that the second-order terms
might be significant, the expansion around a non-stationary point is already
questionable.

However, the evolution of a wave packet that is sharply localized in space
can easily be studied numerically. Fig.~1 shows an example for $\sigma = d$,
$k_0 = 0$, and $eF_0d = \Delta/10$. At time $t=0$ the wave packet is localized
around the site $\ell = 0$. The occupation probabilities $|f_{\ell}|^2$
of the lattice sites, indicated by the open circles, have been connected by
lines to guide the eye. The short dashes correspond to $t=0$, the long dashes
to $t=T_{Bloch}/4$, and the full line represents the wave packet at
$t = T_{Bloch}/2$. After half a Bloch period, the maximum has moved by 9 sites
to the left, whereas it would be 10 sites under ``semiclassical'' conditions.
But even now the basic features of a Bloch oscillation are visible.
The occupation probabilities at $t = 3 T_{Bloch}/4$ are the same as those
for $t = T_{Bloch}/4$, and at $t = T_{Bloch}$ the wave packet regains its
initial shape. 

Fig.~2 depicts the same scenario for $\sigma = 5d$; all other parameters
are unchanged. This is a Bloch oscillation in the classical sense: the
wave packet simply oscillates in space, with its shape remaining almost
invariant.

If the particle is initially localized entirely at a single site $\ell_0$,
i.e., if $f_{\ell}(0) = \delta_{\ell,\ell_0}$, the time evolution
is given directly by the propagator~(\ref{PRO}):
\begin{equation}
|f_{\ell}(t)|^{2} = \left| J_{\ell - \ell_0}\!\left(\frac{\Delta}{eF_0d}
\sin(eF_0dt/2) \right) \right|^{2}
\end{equation}
In this extreme limiting case the wave function's center remains
unchanged in time, but its width oscillates.

\section{The Effect of Combined Static and Oscillating Fields:
	Dynamic Localization} 

In order to solve the time-dependent Schr\"odinger equation
$i\partial_t\psi(t) = \left[ H_0 + H_{int}(t) \right]\psi(t)$
for the model described by~(\ref{TBS}) and (\ref{HIN}), in the general case
where $H_{int}(t)$ contains both a dc field $F_0$ and an additional
ac field $F_1$, it is helpful to introduce the vector potential    
\begin{equation}
A(t) = -F_0 t - \frac{F_1}{\omega}\sin(\omega t)
\label{VEC}
\end{equation}
and employ the gauge transformation
\begin{equation}
\widetilde{\psi}(t) = e^{-ieA(t)x}\psi(t)	\; ,	
\label{TRA}
\end{equation}
where $x$ denotes the position operator, 
$x = \sum_{\ell} |\ell\rangle \ell d \langle\ell| $. This restores spatial
translational symmetry: the Schr\"odinger
equation for the new wave function $\widetilde{\psi}(t)$ reads
$i\partial_t\widetilde{\psi}(t) = \widetilde{H}(t)\widetilde{\psi}(t)$, where 
\begin{equation}
\widetilde{H}(t) = -\frac{\Delta}{4}\sum_{\ell}\left(
    e^{-ieA(t)d}|\ell+1 \rangle \langle \ell|
    \, + \, |\ell \rangle \langle \ell+1| e^{ieA(t)d} \right) \; . 
\label{HTR}
\end{equation}
Then the wave vector is a good quantum number in this gauge, even in the
presence of the electric field.
Observing that the Bloch waves $\chi_k$ (see~(\ref{BLO})) obey the equation
\begin{equation}
\widetilde{H}(t)\chi_k = E\left(q_k(t)\right)\chi_k
\end{equation}
with
\begin{equation}
q_k(t) = k - eA(t) \; ,
\end{equation}
one immediately obtains solutions
\begin{equation}
\widetilde{\psi}_k(t) = \sum_{\ell} |\ell\rangle \exp\!\left(-ik\ell d
-i\int_{0}^{t}\! {\mbox d}\tau \, E(q_{k}(\tau))\right)	\; .
\label{SOL}
\end{equation}
Inverting the transformation~(\ref{TRA}), one finds the wave functions	
\begin{equation}
\psi_{k}(t) = \sum_{\ell} |\ell\rangle \exp\!\left(-iq_{k}(t)\ell d
-i\int_{0}^{t}\! {\mbox d}\tau \, E(q_{k}(\tau))\right) \; .
\label{HOU}
\end{equation} 
These are precisely the Houston states~\cite{Houston40} for the
driven tight-binding lattice. 
 
How are these states related to the Floquet states of the problem?
After all, the Hamiltonian $H_0 + H_{int}(t)$ is $T$-periodic, with
$T = 2\pi/\omega$, so it might appear natural to search for $T$-periodic
solutions $u(t)$ to the eigenvalue equation
$[H_0 + H_{int}(t) - i\partial_t]u(t) = \varepsilon u(t)$.
However, it must be recognized that the problem now actually contains
{\em two} frequencies, the ac frequency $\omega$ and the Bloch frequency
$\omega_{Bloch} = eF_0d$. As a consequence, the transformed Hamiltonian
$\widetilde{H}(t)$ (see~(\ref{HTR})) is, in general, {\em not} $T$-periodic.
Since $\widetilde{H}(t)$ contains both frequencies on an equal footing, 
one can then employ many-mode Floquet theory~\cite{Chu89}.
If $\omega_{Bloch}/\omega$ is a rational number, 
$\omega_{Bloch}/\omega = p/q$ with $p,q$ relatively prime integers, then
$\widetilde{H}(t)$ is periodic in time with period
$\widetilde{T} = q \cdot 2\pi/\omega = p \cdot 2\pi/\omega_{Bloch}$;
otherwise $\widetilde{H}(t)$ is quasiperiodic. 
The Bessel function expansion~(\ref{GEN}) readily yields
\begin{equation}
E(q_k(\tau)) = -\frac{\Delta}{2} \sum_{r = -\infty}^{+\infty}
J_r\!\left(\frac{eF_1 d}{\omega}\right)\cos(kd + [eF_0d + r\omega]\tau) \; . 
\label{BES}
\end{equation}
If there is no integer $n$ such that $eF_0d = n\omega$, then the time integral
over $E(q_k(\tau))$ is an oscillating function with frequencies that are
either multiples of the ac frequency $\omega$ or of the Bloch frequency.
In this case the wave functions~(\ref{SOL}) are two-mode Floquet states
to the Hamiltonian~(\ref{HTR}). But this purely formal discussion is of
limited physical value, since all quasienergies coincide (they are all zero,
up to integer multiples of the respective frequencies), and do not
depend on the field parameters. 

The situation is entirely different if $\omega_{Bloch}/\omega = n$ for
some integer $n$, --- i.e., if
\begin{equation}
e F_0 d = n \omega  \; ,
\label{REF}
\end{equation}
so that the ac field is tuned to an ``$n$-photon-resonance'' with the
Wannier-Stark ladder. Then $\widetilde{H}(t)$ is $T$-periodic, as is
$H_0 + H_{int}(t)$, so there are $T$-periodic Floquet states.
Now the expansion~(\ref{BES}) contains the zero frequency term $r = -n$,
which gives a linearly growing contribution to the total phase after time
integration. But all other terms in the expansion of
$\int_0^t \! {\mbox d}\tau \, E(q_k(\tau))$ are $T$-periodic.
Hence, if~(\ref{REF}) is satisfied, the quasienergies of the Floquet
states~(\ref{SOL}) are determined by the average growth rates (the zero
frequency parts of~(\ref{BES})) of the phases~\cite{HolthausHone93,Zak93}:
\begin{eqnarray}
\varepsilon(k) & = & \frac{1}{T}\int_{0}^{T}\! {\mbox d}t \, E(q_k(t))
\qquad \bmod \omega     \nonumber   \\
& = & (-1)^{n} J_{n}\!\left(\frac{eF_1 d}{\omega}\right) E(k)
\qquad \bmod \omega     \; .
\label{EPK}
\end{eqnarray}
The gauge transformation~(\ref{TRA}) is a $T$-periodic operation only
if~(\ref{REF}) is satisfied. Therefore, the Houston states~(\ref{HOU})
coincide with the Floquet states to $H_0 + H_{int}(t)$ {\em only} in this
resonant case, $\psi_k(t) = u_k(t)\exp(-i\varepsilon(k)t)$. The
quasimomentum $k$ then remains a good quantum number for the eigenfunctions
$u_k(t)$, despite the presence of the dc field, so that~(\ref{EPK}) should be
regarded as a quasienergy-quasimomentum dispersion relation. The particular
significance of the resonance condition~(\ref{REF}) has been discussed
by Zak~\cite{Zak93} with emphasis on commuting time translations and
electric translations. In the following, we will focus entirely on the
resonant case.

Whereas the Wannier-Stark energy eigenstates~(\ref{WAN}) pertaining to dc
fields are {\em localized}, an additional resonant ac field leads to
{\em extended} quasienergy eigenstates that form quasienergy bands of finite
width. Therfore, a resonant ac field can destroy Wannier-Stark
localization, even if its amplitude is infinitesimally small. An arbitrary,
initially localized wave packet $\psi(t=0)$ can be represented as a
superposition of Floquet states with a certain momentum distribution
$\widehat{f}(k,0)$, and the time evolution of each component is simply
determined by its quasienergy phase factor:
\begin{equation}
\widehat{f}(k,T) = \widehat{f}(k,0)\exp(-i\varepsilon(k)T)  \; .
\label{POT}
\end{equation} 
Assuming that the initial momentum distribution is sharply centered around
$k_0$, the time-averaged group velocity of the wave packet is 
\begin{equation}
\bar{v} = \left.\frac{{\mbox d}\varepsilon(k)}{{\mbox d}k}\right|_{k_0}
    = (-1)^n \frac{\Delta d}{2} J_n\!\left(\frac{eF_1 d}{\omega}\right)
    \sin(k_0d)	\; .
\end{equation}
Generally, therefore, the wave packet moves over the whole lattice,
simultaneously spreading. Assuming the initial wave packet to be a
Gaussian~(\ref{GAU}) with width $\sigma_0$, the width $\sigma(sT)$ after
$s$ cycles is given, within the Gaussian approximation to the integrals
defining the spatial Fourier transformation, by
\begin{equation}
\sigma(sT) = \sigma_0\left(1 + \left[sT\frac{\Delta d^2}{4\sigma_0^{2}}
    J_n\!\left(\frac{eF_1d}{\omega}\right)\cos(k_0d)\right]^2\right)^{1/2}
    \; .
\label{NDI}
\end{equation}
Remarkably, the spreading of the wave packet is substantially hindered
if $k_0 = \pm\pi /2d$, so that $\cos(k_0d)$ vanishes. The remaining slow
spreading of the packet then is entirely related to terms beyond the Gaussian
approximation.    

In the special case when $e F_1 d/\omega$ coincides with a
zero $j_{n,s}$ of the Bessel function $J_n$, all quasienergies
$\varepsilon(k)$ coincide. In that case $|\psi(t)|^2$ is strictly periodic
in time with period $T= 2\pi/\omega$: on the average, the wave packet
neither moves nor spreads. This has been called ``dynamic localization''
\cite{DunlapKenkre86,Zhao91,ShonNazareno92,IgnatovEtAl95,MeierEtAl95b}.
In view of the ``collapse'' of the quasienergy band~\cite{Holthaus92b} at the
zeros of $J_n$, dynamic localization can be interpreted as prohibited dephasing:
according to~(\ref{POT}), all components of the wave function acquire exactly
the same phase factor during one period $T = 2\pi/\omega$, so that the wave
packet reassembles itself after every period, apart from an unimportant
overall phase factor~\cite{HolthausHone93}.
 
There is a close correspondence between the approximate formula~(\ref{SHI})
for the quasi\-energies of the two-level system and its counterpart~(\ref{EPK})
for the infinite lattice: if the centers of the two wells are separated by
a distance $d$, then the dipole $\mu$ is approximately $ed/2$, so that
the argument of the Bessel function in~(\ref{SHI}) becomes $eF_1d/\omega$,
as for the infinite lattice. But whereas the approximate formula~(\ref{SHI})
for the two-level system, or the formula~(\ref{REG}) for the
``renormalization'' of the atomic $g$-factor, is a perturbative result,
the corresponding formula~(\ref{EPK}) for the quasienergy band is exact.
This is due to exact translational symmetry of the infinite lattice.
It is quite remarkable that the modification of the energy level structure
by a Bessel function, which appears quite naturally in the context of
infinite lattices, still remains a good approximation for finite lattices,
and even survives in the limiting case of a ``lattice'' with only
two sites~\cite{Holthaus92c}. 

If one formulates the problem of the interaction of particles in spatially
periodic potentials with temporally periodic fields in terms of Floquet states  
and quasienergy bands, one is adopting a point of view quite similar to
the ``dressed atom''-picture~\cite{CohenTannoudji68,CohenTannoudji92}
in atomic physics: the ac field is not regarded as a perturbation. Rather, the
lattice and the ac field together form a new object, the ``dressed lattice'',
and this object has its own quasienergy bands. Exactly as an ac field can
``renormalize'' an atomic $g$-factor, it can also ``renormalize'' the
energy band structure of a lattice. In the following sections we will show
that this point of view is particularly useful for understanding effects
caused by lattice imperfections in the presence of strong ac fields.

\section{Isolated Defects: Manipulating Localization Lengths by ac Fields}

Now we introduce an isolated defect into the lattice and study the
dynamics governed by the Hamiltonian,
$H(t) = H_0 + H_{int}(t) + H_{defect}$, where the defect Hamiltonian
\begin{equation}
H_{defect} = \nu_{0} | 0 \rangle \langle 0 |
\label{HDE}
\end{equation}
changes the energy of the atomic state $|0\rangle$ by $\nu_0$.
$H_0$ and $H_{int}(t)$ are again given by~(\ref{TBS}) and (\ref{HIN}),
respectively. We assume that the ac field is resonant, $eF_0 d = n\omega$,
and investigate the behavior of the Floquet state localized around the defect.  

If we were to consider the simpler problem of a time-independent perturbation
$V$ of the lattice Hamiltonian $H_0$, we could employ the usual resolvent
operator formalism: the energy eigenvalues of the perturbed Hamiltonian
$H = H_0 + V$ are given by the poles of the resolvent
\begin{equation}
R(z) = \frac{1}{z - H} = R^{(0)}(z) + R^{(0)}(z) V R(z)
\label{RES}
\end{equation}
in the complex $z$-plane, where
\begin{equation}
R^{(0)}(z) = \frac{1}{z - H_0}
\end{equation}
is the resolvent of the unperturbed system. In a representation of
spatially localized states the resolvent equation~(\ref{RES}) becomes
the Dyson equation
\begin{equation}
R_{j\ell}(z) = R^{(0)}_{j\ell}(z)
+ \sum_{j'\ell'} R^{(0)}_{jj'}(z) V_{j'\ell'}  R_{\ell'\ell}(z) \; .
\label{DYS}
\end{equation}
If the perturbation $V$ has only a finite number of nonvanishing matrix
elements $V_{j'\ell'}$, then the right-hand side of~(\ref{DYS}) involves
$R_{\ell'\ell}(z)$ only for a finite number of sites $\ell'$. Therefore,
after~(\ref{DYS}) has been inverted for that subset of resolvent operator
matrix elements, the equation itself gives $R_{j\ell}(z)$ for {\em all}
sites $j$.   

The task now is to extend this formalism to the case where there is an
additional time-periodic external field~\cite{HoneHolthaus93}.
To this end we regard the lattice together with the driving field as the
unperturbed system; the perturbation is just the defect potential.
In order to solve the eigenvalue equation~(\ref{EIG}), we have to compute
the resolvent of the quasienergy operator
\begin{equation}
{\cal H} \equiv H(t) - i\partial_t
\end{equation}
in the extended Hilbert space of $T$-periodic functions.

A complete set of spatially localized states
in that space is given by the functions
\begin{equation}
|\ell,m\rangle = |\ell\rangle \exp(ieA(t)\ell d + im\omega t) 
\label{SLS}
\end{equation}
with the vector potential $A(t)$ (see~(\ref{VEC})). With respect to
this set we define Green's functions
\begin{equation}
G_{j\ell}(n,m) = \langle\langle j,n | \frac{1}{z - {\cal{H}}} | \ell,m
    \rangle\rangle  \; , 
\end{equation}
where the double brackets indicate the scalar product~(\ref{SCA}). We no
longer explicitly indicate the dependence of these Green's functions on $z$.
The Dyson equation becomes
\begin{equation}
G_{j\ell}(n,m) = G_{j\ell}^{(0)}(n,m)
    + \sum_{j'\ell'n'}G_{jj'}^{(0)}(n,n') \, V_{j'\ell'} \,
    G_{\ell'\ell}(n',m)    \; ,
\label{QDY}
\end{equation}
where $G_{j\ell}^{(0)}(n,m)$ are the unperturbed Green's functions pertaining
to the operator $(H_0 + H_{int}(t) - i\partial_t)$, which does not include
the time-independent defect potential $V$. Compared to~(\ref{DYS}), the
summation now also includes the additional Fourier index $n'$.

In order to compute the unperturbed Green's functions, we employ the
spatially extended quasienergy eigenfunctions (with integer $s$)
\begin{eqnarray}
|k,s\} & = & u_k(t)e^{is\omega t}	\nonumber   \\
& = & \sum_{\ell} |\ell\rangle \exp\left(-iq_k(t)\ell d
-i\int_{0}^{t}\! {\mbox d}\tau \, [E(q_k(\tau)) - \varepsilon(k)]
+ is\omega t\right) \,
\end{eqnarray}
constructed from the Houston states~(\ref{HOU}). Curly brackets are used to
distinguish these eigenfunctions from the localized functions~(\ref{SLS}).
We then have
\begin{equation}
G_{j\ell}^{(0)}(n,m) = \sum_{k,s} \langle\langle j,n | k,s \}\}
    \frac{1}{z - \varepsilon(k) - s\omega}
    \{\{ k,s | \ell,m \rangle\rangle	\; .
\end{equation} 
The matrix elements can now be expressed as 
\begin{equation}
\langle\langle j,n | k,s \}\} = e^{-ikjd} F_{s-n} \; ,
\end{equation}
where 
\begin{equation}
F_{s-n} = \frac{1}{T}\int_{0}^{T} \! {\mbox d}t \, e^{i(s-n)\omega t} 
\exp\left(-i\int_{0}^{t}\! {\mbox d}\tau \,
[E(q_k(\tau)) - \varepsilon(k)]\right)	\; .
\label{FSN}
\end{equation}
Obviously, the formalism considered here can be of practical use only
if it is possible to neglect all but a finite, small number of Fourier
components. Such an approximation is justified in the case of high
frequencies: the expansion (cf.~(\ref{BES}))
\begin{equation}
\int_{0}^{t} \! {\mbox d}\tau [E(q_k(\tau)) - \varepsilon(k)] =
-\frac{\Delta}{2}\sum_{r\ne-n}J_r\!\left(\frac{eF_1d}{\omega}\right)
\frac{\sin(kd + (n+r)\omega t) - \sin(kd)}{(n+r)\omega}
\end{equation}
shows that the integral on the left hand side is --- at most --- of order
$\Delta/\omega$; the oscillatory character of the function $E(k)$ makes it,
in general, much smaller than this upper limit, as shown by the explicit
expression on the right hand side. Therefore, for frequencies $\omega$ larger
than the bandwidth $\Delta$ a reasonable approximation to~(\ref{FSN}) is
\begin{equation}
F_{s-n} \approx \delta_{s,n}  \quad ,
\label{APR}
\end{equation}
and the unperturbed Green's functions become  
\begin{equation}
G_{j\ell}^{(0)}(n,m) \approx \frac{d}{2\pi}\int_{-\pi/d}^{\pi/d}\! {\mbox d}k \,
    \frac{e^{ikd(\ell-j)}}{z - \varepsilon(k) - n\omega}\delta_{n,m} \; .
\end{equation}
In this high-frequency limit the Green's functions are diagonal in the
Fourier index. In the following we will restrict ourselves to this case,
and suppress the redundant Fourier index altogether.

After inserting the quasienergy-quasimomentum dispersion relation  
\begin{equation}
\varepsilon(k) \; = \; (-1)^{(n+1)} J_{n}\!\left(\frac{e F_1 d}{\omega}\right)
    \frac{\Delta}{2} \cos(kd) \; \equiv \; \frac{W}{2}\cos(kd) \; ,
\end{equation}
where 
\begin{equation}
W = (-1)^{(n+1)} J_{n}\!\left(\frac{e F_1 d}{\omega}\right) \Delta 
\end{equation}
denotes the width of the quasienergy band (up to a sign),
the unperturbed Green's functions can be calculated with the help of
the residue theorem:
\begin{equation}
G_{j\ell}^{(0)} = \frac{4}{W}
    \frac{\zeta_0^{|\ell - j| + 1}}{1 - \zeta_0^2}  \; ,
\end{equation}
where
\begin{equation}
\zeta_0 = \frac{2z}{W} \pm \sqrt{\frac{4z^2}{W^2} - 1}	\; .
\end{equation}
The sign has to be chosen such that $\zeta_0$ falls inside the unit circle.

For an isolated defect of the type given by~(\ref{HDE}) we now have
\begin{equation}
V_{j\ell} = \nu_0\delta_{j,0}\delta_{\ell,0}  \; ,
\end{equation}
and the Dyson equation~(\ref{QDY}) becomes in the high-frequency limit
\begin{equation}
G_{j\ell} = G_{j\ell}^{(0)} + \nu_0 G_{j0}^{(0)} G_{0\ell}	\; .
\end{equation}
Specifying $j=0$, solving for $G_{0\ell}$, and reinserting, one obtains 
\begin{equation}
G_{j\ell} = G_{j\ell}^{(0)}
    + \frac{\nu_0 G_{j0}^{(0)} G_{0\ell}^{(0)}}{1 - \nu_0 G_{00}^{(0)}} \; .
\end{equation}
The pole which lies inside the unit circle, $|\zeta_0| < 1$, is found to be
\begin{equation}
\zeta_0 = \mbox{sign}(\nu_0/W)\left(\sqrt{\frac{4\nu_0^2}{W^2} + 1}
    - \left|\frac{2\nu_0}{W}\right|\right)    \; .
\end{equation}
Hence, the quasienergy of the state supported by the defect is given by
\begin{equation}
\varepsilon_{defect}
    \; = \; \frac{W}{4}\left(\zeta_0 + \frac{1}{\zeta_0}\right)
    \; = \; \mbox{sign}(\nu_0)\sqrt{\nu_0^2 + \frac{W^2}{4}}	\; .
\label{DEF}
\end{equation}
Thus, while the width $W$ of the quasienergy band oscillates when the
parameter $eF_1d/\omega$ is varied, the quasienergy of the defect state
follows these oscillations, maintaining a positive distance from the band.

The oscillations of the band width have a pronounced effect on the spatial
density of the defect state. In the high-frequency regime, the probability
$p(\ell)$ for the defect state to occupy the $\ell$-th site is determined
by the residue of the diagonal Green's function $G_{\ell\ell}$ in the
complex quasienergy plane~\cite{HoneHolthaus93}. An elementary calculation
yields
\begin{equation}
p(\ell) = \frac{|\nu_0|}{\sqrt{\nu_0^2 + W^2/4}}
    \left(\sqrt{\frac{4\nu_0^2}{W^2} + 1}
    - \left|\frac{2\nu_0}{W}\right|\right)^{2|\ell|}	\; .
\label{ANA}
\end{equation}
Thus, the occupation probability falls off exponentially from the site
$\ell =0$ where the defect is located. The inverse exponential decay length,
measured in multiples of the lattice spacing $d$, is given by 
\begin{equation}
\left(\frac{L}{d}\right)^{-1} = -2\ln\left(\sqrt{\frac{4\nu_0^2}{W^2} + 1}
    - \left|\frac{2\nu_0}{W}\right|\right)    \; .
\end{equation}
Obviously, $L^{-1}$ increases monotonically with $|\nu_0/W|$, the ratio
of the size of the perturbation and the quasienergy band width. Therefore,
it is possible to manipulate the spatial extension of defect states by
tuning the parameters of the driving field. When $eF_1d/\omega$ approaches
a zero of the Bessel function $J_n$, while $eF_0d = n\omega$, the defect
state should be concentrated wholly at the anomalous
site~\cite{HoneHolthaus93}. This possibility of controlling localization
lengths by external fields is one of the most interesting predictions of the
``dressed lattice''-picture.  

To check the validity of our conclusions even away from the high frequency
limit, we perform numerical computations for a finite lattice with $N = 101$
sites, with a defect placed in the center. The wave functions vanish at the
chain ends. Fig.~3 shows the quasienergy spectrum for $\omega/\Delta = 1.0$ and
$eF_0 d = \omega$, i.e., for a ``one-photon resonance''. The defect strength
is $\nu_0/\Delta = 0.1$. The quasienergy band structure is determined by the
Bessel function $J_1$; the ``collapse'' of the band at $j_{1,1} = 3.83171$
and $j_{1,2} = 7.01559$ is indicated by the arrows. As expected, the
quasienergy of the defect state on top of the band follows the oscillations
of the band width.

To investigate the degree of localization of the defect state, we compute
its quasienergy eigenfunction
\begin{equation}
u_{defect}(t) = \sum_{\ell = 1}^{N} c_{\ell}^{(defect)}(t) |\ell\rangle
\end{equation}
and calculate the time-dependent inverse participation ratio
\begin{equation}
P_{defect}(t) = \sum_{\ell = 1}^{N} | c_{\ell}^{(defect)}(t)|^{4} \; .
\label{PDE}
\end{equation}
If the defect state were uniformly extended over all $N$ sites,
$P_{defect}(t)$ would vanish as $1/N$, whereas it approaches unity for a
completely localized state. Fig.~4 shows numerical results for the same
situation as considered in Fig.~3. The values of the scaled ac field
strength $eF_1d/\omega$ are $j_{1,1}$ (short dashes), 5.0 (long dashes),
and $j_{1,2}$ (full line). The density of the defect state exhibits a periodic
oscillation that leads to the observed time-dependene of the inverse
participation ratio. These oscillations become negligibly small for higher
frequencies; $P_{defect}$ then becomes independent of time. But clearly,
even for $\omega = \Delta$ the localization of the defect state is much
more pronounced at the zeros of the Bessel function than at other parameters.

In order to test the prediction of~(\ref{ANA}) directly against the numerical
data, we compute the participation ratio as a function of $eF_1d/\omega$ at
$t = 3T/4$, when the ac field vanishes. The result is shown as the full
line in Fig.~5. The broken line, on the other hand, results from the
evaluation of the approximate formula~(\ref{ANA}):
$P_{defect} \approx \sum_{\ell} (p(\ell))^2$.
The two lines differ for low field strengths, as the approximate expression
goes to unity when $F_1$ vanishes, whereas the exact result approaches a
smaller limit, representing a state spread over more than the single impurity
site. However, the agreement is strikingly
good for large ac amplitudes. We conclude that the high-frequency
approximation can give good results even if the frequency $\omega$ is not
really large compared to the bandwidth $\Delta$.    

It should be stressed that in the high-frequency regime $\omega \ge \Delta$
where the approximation~(\ref{APR}) is valid the effect of the driving field
is fully described by the quasienergy spectrum: the {\em driven} lattice with
{\em quasienergy} band width $W$ becomes equivalent to an {\em undriven}
lattice with that same {\em energy} band width. This observation stets the
stage for the following discussion of Anderson-type localization in the
presence of ac fields.

\section{Random Disorder: Controlling Anderson Localization}

Having dealt with the effects of a single defect, we now
consider randomly disordered lattices:
\begin{equation}
H(t) = H_{0} + H_{int}(t) + H_{random}
\end{equation}
where $H_{0}$ and $H_{int}(t)$ are given by~(\ref{TBS}) and (\ref{HIN}),
as before, and $H_{random}$ introduces site-diagonal disorder:
\begin{equation}
H_{random} = \sum_{\ell} \nu_{\ell} \, |\ell\rangle\langle\ell| \; .
\end{equation}
For specificity we take the probability distribution $\rho(\nu)$ for the
random energies $\nu_{\ell}$ as  
\begin{equation}
\rho(\nu) = \frac{1}{\pi\nu_{max}}
    \frac{1}{\sqrt{1 - \left(\frac{\nu}{\nu_{max}}\right)^2}}	
\label{DIS}
\end{equation}
for $|\nu| < \nu_{max}$, and zero otherwise, though the precise choice of
$\rho(\nu)$ does not greatly affect the conclusions below.
Without external fields, the time independent Hamiltonian $H_{0} + H_{random}$
defines a paradigmatic system for the study of Anderson
localization~\cite{Anderson58,MottTwose61,Thouless74,KramerMacKinnon93}.
It admits only localized states, for arbitrary positive disorder strength
$\nu_{max}$. The typical localization length $L$ is given by the square of
the ratio of the bandwidth $\Delta$ and $\nu_{max}$~\cite{Thouless79}:
\begin{equation}
L \sim \left(\frac{\Delta}{\nu_{max}}\right)^2 d    \; .
\end{equation}  
Thus, if the disorder strength becomes comparable to the unperturbed
bandwidth, the extension of typical energy eigenstates is only of the
order of a few sites.

Consequently, the fact that in the presence of resonant high-frequency ac
fields the amplitude-dependent quasienergy band width $W$ takes over the
role of the energy band width $\Delta$ leads to an interesting
prediction~\cite{HolthausEtAl95a}:
It must be possible to manipulate the degree of Anderson localization by
tuning the dimensionless ac field strength $z = e F_1 d/\omega$, provided the
ac field is resonant with the Wannier-Stark ladder, i.e., $e F_0 d = n\omega$
for some integer $n$. To demonstrate this effect, we compute the Floquet
eigenfunctions for a finite disordered lattice with $N = 101$ sites,
\begin{equation}
u_{m}(t) = \sum_{\ell=1}^{N} c_{\ell}^{(m)}(t) \, |\ell\rangle \; ,
\end{equation}
and then determine the average inverse participation ratio $P$:
\begin{equation}
P \equiv \frac{1}{N} \sum_{\ell,m=1}^{N} |c_{\ell}^{(m)}(3T/4)|^{4} \; .
\label{PAV}
\end{equation}
As in the previous case of an isolated defect, the choice of the particular
moment $t = 3T/4$ is motivated by the fact that the ac field vanishes
at that time (see~(\ref{HIN})), but it is without special significance. In the
high-frequency limit, where the time-dependence of the Floquet states becomes
weak, $P$ is almost independent of the particular time at which it is evaluated.

Fig.~6 shows $P$ versus $z = eF_1d/\omega$ for $n = 1$ and $\omega = \Delta$,
with $\nu_{max}/\Delta = 0.01$, $0.05$, $0.10$, and $0.20$. The effect of the
collapse of the quasienergy band is quite dramatic: when $z$ approaches
a zero of $J_1$, the Floquet states are localized almost entirely at individual
sites, as testified by values of $P$ close to unity.

We thus have to distinguish clearly the following effects: A dc field applied
to a tight-binding lattice results in Wannier-Stark localization; every wave
packet can be represented as a linear combination of {\em localized} energy
eigenstates. An additional ac field, tuned exactly to an ``$n$-photon
resonance'' with the Wannier-Stark ladder, leads to {\em extended} quasienergy
eigenstates, and an initially localized wave packet will, in general, move
throughout the lattice and spread, unless $eF_1d/\omega$ equals a zero of $J_n$.
At these special parameters there is no quasienergy-quasimomentum dispersion,
and the wave packet, although built up from extended states, remains
``dynamically'' localized. If there is additional random disorder, the
quasienergy eigenstates exhibit Anderson localization, with amplitude-dependent
localization lengths. Anderson localization is most pronounced at the very
same parameters where there would be dynamic localization in an ideal lattice.
To illustrate the profound difference between dynamic localization and
field-induced Anderson localization, we show in Fig.~7 the quasienergy
spectrum corresponding to the case of strongest disorder considered in Fig.~6,
$\nu_{max}/\Delta = 0.20$.
At the zeros of $J_1$, where the quasienergy band of the ideal lattice
would collapse --- a necessary condition for dynamic
localization~\cite{HolthausHone93} --- the quasienergies of the disordered
lattice are scattered over a range of $2\nu_{max}$. The components
of an initially localized wave packet will, therefore, dephase in the course
of time, in contrast to the ideal case. Nevertheless, the wave packet will
remain localized, since all the Floquet states are individually localized
themselves.
      
In short: a dc field gives rise to localized energy eigenstates, an
additional resonant ac field then leads to extended quasienergy eigenstates,
and random disorder results in Anderson localization of these quasienergy
states. Thus, localization in randomly disordered, resonantly driven
tight-binding lattices has nothing to do with the Wannier-Stark localization
that would result from a dc field applied to an ideal lattice.
  
To substantiate this statement, we confirm the scaling behavior expected for
genuine Anderson-type localization. The relevant dimensionless parameter of
the problem should be the ratio of the quasienergy band width and the disorder
strength. If this hypothesis is correct, then lattices with different
disorder strengths must yield the same average inverse participation ratio $P$,
if their values of $\Delta J_n(e F_1 d/\omega)/\nu_{max}$ coincide. 
In Figs.~8, 9, and~10 we plot $P$ for $n = 1$, $n=2$, and $n = 3$,
respectively, again for lattices with $N = 101$ sites and
$\omega/\Delta = 1.0$.
The disorder strength varies from $\nu_{max}/\Delta = 0.01$ to $0.10$, in
steps of $0.01$. As can be seen, a certain characteristic ac field strength
is necessary to effectively counteract the localization found for vanishing
ac amplitude. As a measure of this characteristic field strength we take
that value $z_{1/2}$ of the dimensionless amplitude $z = eF_1 d/\omega$
for which $P(z)$ is reduced to half the ``Wannier-Stark''-value:
$P(z_{1/2}) = P(0)/2$. Since $P$ should depend only on the ratio
$\Delta J_n(z)/\nu_{max}$, there should be a relation~\cite{HolthausEtAl95b}
\begin{equation}
\Delta J_n(z_{1/2}) = c_n \nu_{max} \; ,
\label{AND}
\end{equation}
with numbers $c_n$ of order unity. The $n$-dependence of these numbers results
from the fact that the degree of localization for $z = 0$ depends on $n$,
i.e., on the strength of the dc field. As long as the characteristic ac
amplitudes remain weak, $z_{1/2} < n/2$, one can approximate the Bessel
functions $J_n$ by their lowest order terms, $J_n(z) \approx (z/2)^n/n!$,
to get
\begin{equation}
z_{1/2} \approx 2\left( n! \, c_n \, \frac{\nu_{max}}{\Delta} \right)^{1/n}
\; .	\label{APJ}
\end{equation}
Provided the disorder is so weak that the characteristic ac amplitudes
also remain sufficiently small, these amplitudes should scale approximately with
the $n$-th root of the disorder strength~\cite{HolthausEtAl95b}.

Fig.~11 confirms these hypotheses. The boxes denote
the characteristic amplitudes for $n = 1$, $2$, and $3$, as determined from
the preceding three figures. The dashed lines follow from the approximate
relation~(\ref{APJ}), with $c_1 = 1.074$, $c_2 = 1.183$, and $c_3 = 1.189$.
The agreement with the numerical data points is very good for $n=1$ and
$n = 2$, though there are clear deviations for $n=3$. But these deviations 
merely indicate that the lowest-order approximation to $J_3$ is already
insufficient: if the full Anderson relation~(\ref{AND}) is evaluated without
that approximation one obtains the solid line, which gives a truly excellent
fit to the numerical data. This impressive agreement between the numerical
results and the Anderson relation~(\ref{AND}) confirms that we are not
dealing with remnants of Wannier-Stark localization induced by the dc field,
but rather with Anderson localization resulting solely from disorder.  

>From a dynamical point of view, it is of interest to study directly the effect
that Anderson localization of quasienergy states has on the time evolution of
an initially localized wave packet. To this end, we propagate a wave packet
given at $t=0$ by a Gaussian~(\ref{GAU}) in a lattice with $N = 1001$ sites.
The field parameters are $\omega/\Delta = 1.0$, $eF_0d/\omega = 1.0$,
and $eF_1d/\omega = 0.3$. The diagonal disorder is again
distributed according to~(\ref{DIS}), with $\nu_{max}/\Delta = 0.1$.
The initial width of the packet is $\sigma = 5d$, and the initial
momentum $k_0 = \pi/(4d)$.
Fig.~12 depicts the deviation of the wave packet's center,
$\widehat{x} = \sum_{\ell} \ell d |f_{\ell}(t)|^2$, from the initial position
$\widehat{x}_{0} = 0$, whereas Fig.~13 shows the evolution of its width
$\sigma$, both for a period of 1000~cycles $T = 2\pi/\omega$ of the external
field. As expected, the packet does not move on the average, the typical
excursion length being merely one or two lattice constants, and its width does
not increase appreciably. If there were no disorder, the packet would move
by more than 300 lattice sites to the left within 1000 ac cycles, and its width
would grow as indicated by the dashed line in Fig.~13.

As in the case of an isolated defect, the identification of a periodically
driven lattice of bandwidth $\Delta$ with an undriven lattice of
``renormalized'' bandwidth $\Delta J_n(eF_1d/\omega)$ is essentially correct
in the high-frequency regime, $\omega/\Delta \ge 1$. Nevertheless, localization
effects can efficiently be manipulated by ac fields even if this condition is
not satisfied, provided the dc field alone is strong enough to cause
substantial Wannier-Stark localization. Fig.~14 shows an example for
$\omega/\Delta = 0.1$ and $eF_0 d = 10 \, \omega$, i.e., for a
``10-photon resonance''. The lattice has 101 sites, and the disorder strength is
$\nu_{max}/\Delta = 0.1$. The average inverse participation ratio $P$ remains
almost constant for weak amplitudes, then shows a remarkably sharp crossover
from ``strong'' to ``weak'' localization, and exceeds $0.5$ again in the
vicinity of $j_{10,1} \approx 14.5$. At the second positive zero of $J_{10}$,
i.e., at $j_{10,2} \approx 18.4$, localization is less pronounced.

Fig.~15 shows the quasienergy spectrum for this case. Since
$\omega \ll \Delta$, the quasienergy band of the ideal lattice does not fit
into a single quasienergy Brillouin zone, but overlaps itself. Because
of the loss of translational invariance due to disorder, this leads to a dense
mesh of avoided crossings when the quasienergies of the random lattice are
plotted, e.g., versus the ac amplitude. But even though the quasienergy
spectrum appears quite intricate in the low-frequency case, there remains
a clearly visible connection between the behavior of the eigenvalues and that
of the inverse participation ratio.

\section{Interband Transitions}

The discussion so far has completely neglected transitions between
different energy bands. But even if there is only a dc field, interband
transitions can give rise to important and interesting changes of the
dynamics~\cite{Zener34,RotvigEtAl95,HoneZhao95}. In order to study the
role of interband transitions in the presence of combined ac and dc fields,
we now investigate the two-band model defined by the
Hamiltonian~\cite{FukuyamaEtAl73}
\begin{equation}
H(t) = H^{(1)}(t) + H^{(2)}(t) + H_{interband}(t)	\; .
\label{TBA}
\end{equation}
For either $j=1$ or $j=2$ the Hamiltonian $H^{(j)}(t)$ describes a particle in a
driven tight-binding lattice with a single band of width $\Delta_j$:
\begin{eqnarray}
H^{(j)}(t) & = & (-1)^j\frac{D}{2}\sum_{\ell} |\ell,j\rangle \langle\ell,j|
    + e F(t) d \sum_{\ell} |\ell,j\rangle \ell \langle\ell,j| \nonumber \\ 
& & + (-1)^j \frac{\Delta_{j}}{4}\sum_{\ell}\left(
    |\ell+1,j\rangle \langle\ell,j| + |\ell,j\rangle \langle\ell+1,j|\right) 
    \; .
\label{HHJ}
\end{eqnarray}
$F(t)$ is the total external field,
\begin{equation}
F(t) = F_0 + F_1\cos(\omega t)	\; .
\end{equation}
If both $F_0$ and $F_1$ vanish, then $D$ denotes the energy difference
between the band centers. The interaction between the two bands is modelled by
\begin{equation}
H_{interband}(t) = eF(t)X_{1,2} \sum_{\ell}\left( |\ell,1\rangle \langle\ell,2|
    + |\ell,2\rangle \langle\ell,1| \right) \; .
\end{equation}
Here $X_{1,2}$ is the matrix element of position $x$ between the Wannier
states of the two bands centered on a single site. If we approximate it by
the dipole matrix element between the lowest two energy eigenstates of a
particle in a box of width $d$, as in~\cite{RotvigEtAl95}, we have
$X_{1,2} = -16d/(9\pi^2)$. We neglect, as usual, dipole matrix elements
between Wannier states centered on different sites.

If the ac field is resonant with each of the two interpenetrating Wannier-Stark
ladders originating from the two bands~\cite{FukuyamaEtAl73},
$eF_0d = n\omega$, then there are quasienergy bands with finite widths.
For vanishing interband interaction, $X_{1,2} = 0$, the normalized quasienergy
eigenfunctions for a lattice with $N$ sites are
(neglecting finite size effects)
\begin{equation}
u_k^{(j)}(t) = \frac{1}{\sqrt N}\sum_{\ell} |\ell,j\rangle
\exp\left(-iq_k(t)\ell d - i\!\int_{0}^{t} \! {\mbox d}\tau \,
[E^{(j)}(q_k(\tau)) -\varepsilon^{(j)}(k)] \right)  \; ,
\end{equation}
as in~(\ref{HOU}), where
\begin{equation}
E^{(j)}(k) = (-1)^j\frac{D}{2} + (-1)^j\frac{\Delta_j}{2}\cos(kd)
\end{equation}
and
\begin{equation}
\varepsilon^{(j)}(k) = (-1)^j\frac{D}{2} + (-1)^{(j+n)}\frac{\Delta_j}{2}
J_{n}\!\left(\frac{eF_1d}{\omega}\right)\cos(kd)
\end{equation}
are the energy and quasienergy eigenvalues, respectively.

For nonzero $X_{1,2}$, and resonant ac fields, the two-band model~(\ref{TBA})
is equivalent to a system of two-level systems labelled by the wave vector $k$.
If $(\Delta_1 + \Delta_2)/2 \le \omega$, which will be fulfilled in many
situations of interest, a high-frequency approximation analogous
to~(\ref{APR}) yields the matrix elements ($r$, $s$ integer)
\begin{equation}
\langle\langle u^{(1)}_k(t) e^{ir\omega t}| H_{interband}(t) |
u^{(2)}_{k'}(t) e^{is\omega t} \rangle\rangle \approx
eF_0 X_{1,2} \delta_{r,s} \delta_{k,k'} + \frac{1}{2} eF_1 X_{1,2}
\left(\delta_{r,s+1} + \delta_{r,s-1}\right)\delta_{k,k'}   \; .
\end{equation}
The discussion of section~2 then shows that, besides the resonance
$eF_0d = n\omega$ required for the emergence of quasienergy bands, there is
a second, less obvious resonance: if also
\begin{equation}
2eF_0X_{1,2} = m\omega
\label{RCO}
\end{equation}
for some integer $m$, then the quasienergy bands will be most strongly
affected by $H_{interband}(t)$.

One can also arrive at this second resonance condition~(\ref{RCO}) by a
different argument. If both $\Delta_1 = 0$ and $\Delta_2 = 0$,
then~(\ref{TBA}) reduces to a system of uncoupled two-level systems
$H^{(\ell)}_{tls}(t)$ that are labelled by the site index $\ell$:
\begin{equation}
H^{(\ell)}_{tls}(t) = \frac{D}{2}\sigma_z + eF(t)\ell d \, {\bf 1}
+ eF(t)X_{1,2} \sigma_{x}  \; .
\label{TBZ}
\end{equation}
All of these are unitarily equivalent to $H^{(0)}_{tls}(t)$, which is
just~(\ref{TLS}) with $\Delta$ and $\mu$ replaced by $D$ and $eX_{1,2}$,
respectively. Thus, if~(\ref{RCO}) holds, the quasienergies are determined
approximately by~(\ref{SHI}) for strong fields or high frequencies, for each
$\ell$. For nonvanishing $\Delta_j$ this $\ell$-degeneracy is removed, and
the quasienergy bands obtain finite widths.

We illustrate these qualitative considerations by numerical computations.
First, Fig.~16 shows the quasienergies for the two-level systems~(\ref{TBZ})
with $eF_0d = \omega$ and $X_{1,2} = -16d/(9\pi^2)$, so that~(\ref{RCO}) is
not satisfied. The energy level separation is $D = 5.5 \, \omega$, hence
$\omega/D < 1$. The behavior of the quasienergies in this low-frequency case,
including the appearance of the sharp avoided crossing at
$eF_1d/\omega \approx 17$, can be understood with the help of the arguments
given in ref.~\cite{HolthausHone93}.

Fig.~17 then depicts the quasienergy spectrum for the two-band
model~(\ref{TBA}) with the same parameters $D/\omega$, $eF_0d/\omega$, and
$X_{1,2}/d$. The number of sites is $N = 20$; the bandwidths are
$\Delta_1 = 1.2 \, \omega$ and $\Delta_2 = \omega$. The centers of the two
quasienergy bands coincide precisely with the quasienergies of the associated
two-level system, cf.~Fig.~16. Since we have a ``one-photon-resonance'', the
oscillations of the bandwidths are determined by $J_1(eF_1d/\omega)$. Arrows
on top of the figure indicate values of the scaled ac amplitude $eF_1d/\omega$
that coincide with a zero of $J_1$. At these zeros, both bands collapse. In
addition, there is strong level repulsion at the borders
($\varepsilon/\omega = \pm 1/2$) and in the center ($\varepsilon/\omega = 0$)
of the quasienergy Brillouin zone.

When the interband separation $D$ is decreased, the dynamics become more
complicated. Fig.~18 shows quasienergies for $D = 1.6 \, \omega$; all other
parameters are as in Fig.~17. Some of the band collapses now become
imperfect, similar to the observations in ref.~\cite{RotvigEtAl95}, and
additional band narrowings occur.

In order to satisfy the second resonance condition~(\ref{RCO}) together with
the basic condition $eF_0d = n\omega$, it is necessary that $2 X_{1,2}/d = m/n$
be a ratio of two integers. Of course, it is always possible to find rational
approximants $m/n$ arbitrarily close to any given value of $2 X_{1,2}/d$,
but the most relevant cases are those where both $n$ and $m$ are small.  
Fig.~19 depicts the quasienergy bands for $eF_1d = \omega$ and
$2 X_{1,2}/d = -1$. The other parameters are $D = 3.5 \, \omega$,
$\Delta_1 = 1.2 \, \omega$, and $\Delta_2 = \omega$. Again, arrows on top of
the figure indicate zeros of $J_1$. As expected, the quasienergy spectrum now
appears quite different from those shown in Figs.~17 and~18. In particular,
there are no band collapses at high ac amplitudes.

The first avoided quasienergy band crossing that occurs in Fig.~19 in the
center of the Brillouin zone is shown enlarged in Fig.~20. Such avoided
crossings are of great significance: at an avoided crossing, Fourier spectra
of Floquet functions peak strongly around a certain harmonic of the ac
frequency~\cite{HolthausHone93,HolthausHone94}. This observation has led to
the suggestion that a semiconductor superlattice, irradiated by a strong
far-infrared laser field under conditions which correspond to an avoided
crossing of its quasienergy minibands, might emit radiation preferentially at
some odd integer multiple of the driving frequency
$\omega$~\cite{HolthausHone94}. 
 
The merit of the introduction of quasienergy bands is an enormous conceptual  
simplification. If one uses the energy eigenstates, i.e., the unperturbed
Bloch waves, to describe the dynamics of an electron in a lattice under
simultaneous ac and dc bias, one first has to account for Zener tunneling
that occurs in static fields~\cite{Zener34}. Next, the oscillating field leads
to Rabi-type oscillations between the {\em energy} eigenstates. The systematic
investigation of the effects that additional lattice disorder would have then
appears almost hopeless. The Floquet picture, on the other hand, is much more
economical, since it fully incorporates both external fields in the basis of
quasienergy eigenfunctions. The influence of disorder can then be discussed
with the help of the quasienergy spectrum: the degree of localization 
depends on the quasienergy band widths~\cite{DreseHolthaus95}.

\section{Discussion and Outlook}

Let us summarize: The quantum mechanics of particles in periodic potentials
that interact simultaneously with static and with resonant, oscillating
electric fields can most conveniently be studied by introducing quasienergy
states instead of the usual Bloch waves. The widths of the quasienergy bands
depend strongly on the field parameters, and can even approach zero. This
band width renormalization bears close mathematical and conceptual similarity
to the well-studied renormalization of atomic
$g$-factors~\cite{HarocheEtAl70,Chapman70,YabuzakiEtAl72,YabuzakiEtAl74}. 
If there is lattice disorder,
the ratio of disorder strength and quasienergy band width determines the
localization lengths of quasienergy eigenstates, exactly as the ratio
of disorder strength and energy band width determines the localization
of energy eigenstates in the static case. By adjusting the ac amplitude,
it is possible to switch from the regime of ``weak'' localization, where the
localization lengths are well beyond the length of the entire lattice,
to the regime of ``strong'' localization, where the quasienergy states are
essentially localized at individual sites.

We have restricted the discussion to model systems with nearest neighbor
interactions only, but it is not difficult to drop this
restriction~\cite{ZhaoEtAl95}. For example, consider the case where there
are also next nearest neighbor couplings. Then the dispersion relation
of the undriven lattice contains an additional term proportional to
$\cos(2kd)$, so that quasienergy bands of finite width occur
not only for $eF_0d = n\omega$, but also for $eF_0d = n\omega/2$.
The appearance of additional bands reflects an $n$-photon-resonance
between ``atomic'' states that are next nearest neighbors. 

It has already been mentioned that semiconductor superlattices are possible
candidates for an experimental test of the ideas outlined in this paper. 
For such superlattices the nearest neighbor approximation is appropriate, and
there is always some disorder in these artificially grown mesoscopic
systems. It must be emphasized that an electron in a semiconductor
experiences not only the external field, but
also~\cite{Buettiker93,Buettiker94} a field from induced polarization.
However, it can be shown theoretically that dynamic localization in ideal
superlattices should persist even in the presence of Coulomb
interactions~\cite{MeierEtAl94,MeierEtAl95a}, and signatures of dynamic
localization have actually been observed in experiments performed at
UCSB~\cite{FEL3}. Therefore, we may assume that our models, although highly
idealized, still capture a significant part of the physics of real
superlattices. A recent analysis~\cite{MeierEtAl95b} of a realistic
full three-dimensional model of an anisotropic semiconductor, including
exciton formation as a consequence of coulomb interactions, demonstrates that
the dynamic localization phenomenon is accompanied by a change in the
effective dimensionality of the superlattice exciton.

Entirely different examples of particles in periodically driven,
one-dimensional periodic potentials are provided by ultracold atoms in
modulated standing light waves. If atoms of mass $M$ move in a monochromatic
standing light wave with a frequeny that is detuned sufficiently far from
any atomic resonance, and if the standing wave position is periodically
modulated with an amplitude $\Delta L$ and angular frequency $\omega_m$,
then the effective Hamiltonian can be written
as~\cite{MooreEtAl94,BardroffEtAl95,RobinsonEtAl95,LatkaWest95} 
\begin{equation}
H(t) = \frac{p_x^2}{2M}
    - \frac{\Omega}{8}\cos[2k_L(x - \Delta L \sin(\omega_m t))]	\, ,
\end{equation}
where $\Omega$ denotes the effective Rabi frequency, and $k_L$ is the light
wave number. This Hamiltonian is unitarily equivalent, and quasienergetically
isospectral, to 
\begin{equation}
\widetilde{H}(t) = \frac{p_x^2}{2M} - \frac{\Omega}{8}\cos(2 k_L x)
    -\lambda x \sin(\omega_m t)
    -\frac{\lambda^2}{4M\omega_m^2}	\; ,
\end{equation}
with
\begin{equation}
\lambda = M\omega_m^2 \Delta L	\; .
\end{equation}
The corresponding classical Hamiltonian system exhibits chaotic dynamics.
The above model system has been used to study, both experimentally and
theoretically, the quantum mechanical suppression of classical phase space
diffusion~\cite{MooreEtAl94,BardroffEtAl95,RobinsonEtAl95,LatkaWest95}.
(Incidentally, this suppression of classical diffusion has also been termed
``dynamical localization''. It must be recognized, however, that it is in
no way related to the ``dynamic localization'' discussed in section~4.) 

Clearly, ultracold atoms in standing light waves are higly attractive
candidates for studying quantum dynamics in periodic potentials, because of
the conceptual simplicity of the experimental set-up and the high degree
of control over the parameters~\cite{MooreEtAl94}. In fact, one can not
only tailor the initial momentum distribution of the atoms at will, but
also turn the periodic potential off easily, thus obtaining access to the
momentum distribution of de Broglie waves that have evolved under the
influence of that potential~\cite{BenDahanEtAl96}. Moreover, there is no
dissipation or scattering from defects, and the interaction between the
neutral atoms can be negelected. These unique features have been exploited
in a very recent beautiful experiment~\cite{BenDahanEtAl96}: it has been
demonstrated that ultracold Cesium atoms prepared in the ground energy band
of the potential created by a standing light wave perform Bloch oscillations
when they are driven by a constant force.
Ben Dahan {\em et al.}~\cite{BenDahanEtAl96} were able to deduce the velocity
of the oscillating atoms as function of time and obtained perfect agreement
with the theoretical prediction, thus visualizing for the first time what
Bloch and Zener had envisioned decades ago.

In the light of these achievements, other possibilities come to mind
readily. If the current experiments with {\em periodically} forced
atoms~\cite{MooreEtAl94,RobinsonEtAl95} can be extended into the
single-band tight-binding regime, it should become feasible to study
dynamic localization (in the sense of section~4) of atomic de Broglie waves
--- i.e., spatial localization of ultracold atoms in standing light
waves, caused by collapses of their quasienergy bands. Another interesting
question is suggested by eq.~(\ref{NDI}): can one transport an atomic wave
packet with suitably prepared initial momentum in the field of a standing
light wave (almost) without dispersion?   
It appears quite likely that the already successful combination of
atomic, optical, and solid state physics will unravel further important
effects.

\vspace{20mm}

{\bf Acknowledgments:} M.H.\ wishes to thank Stig Stenholm for bringing
the experimental and theoretical work on the renormalization of atomic
$g$-factors to his attention. We also acknowledge valuable discussions
with S.J.\ Allen, M.\ B\"uttiker, Y.\ Castin, F.\ Gebhard,
S.\ Grossmann, A.-P.\ Jauho, S.W.\ Koch, P.\ Thomas, and X.-G.\ Zhao. 
D.H.\ was supported in part by NSF grant PHY~94-07194.

\begin{figure}
\caption[FIG.~1] {Time evolution of a sharply localized
    wave packet~(\ref{GAU}) in a tight-binding lattice in the presence
    of a static electric field, $eF_0d = \Delta/10.$ The initial localization
    length $\sigma$ is a single lattice period: $\sigma = d$.
    The site populations $|f_{\ell}(t)|^{2}$ are connected by lines to
    guide the eye. Short dashes: $t = 0$, long dashes: $t = T_{Bloch}/4$,
    full line: $t = T_{Bloch}/2$, where $T_{Bloch} = 2\pi/(eF_0d)$ denotes
    the Bloch period.}
\end{figure}

\begin{figure}
\caption[FIG.~2] {Same plots as Fig.~1, but for a wave packet not so sharply
    localized: $\sigma = 5d$.}
\end{figure}

\begin{figure}
\caption[FIG.~3] {Quasienergy spectrum for a finite lattice with $N = 101$
    sites, and an isolated defect in the center. The ac frequency is
    $\omega/\Delta = 1.0$, and $eF_0d = \omega$,
    corresponding to a ``one-photon resonance''. The predicted collapse
    of the quasienergy band at $j_{1,1} = 3.83171$ and $j_{1,2} = 7.01559$
    is indicated by the arrows. The defect strength is $\nu_0/\Delta = 0.1$.
    The quasienergy of the defect state appears above the quasienergy band,
    cf.~(\ref{DEF}).}
\end{figure}

\begin{figure}
\caption[FIG.~4] {Time-dependent inverse participation ratio $P_{defect}(t)$
    (see~(\ref{PDE})) for the same situation as considered in Fig.~3,
    for $eF_1d/\omega = j_{1,1}$ (short dashes), $5.0$ (long dashes),
    and $j_{1,2}$ (full line). The origin of time in this figure, $t=0$,
    corresponds to $t=3T/4$ in the Hamiltonian~(\ref{HIN}), when
    the ac field vanishes.}
\end{figure}

\begin{figure}
\caption[FIG.~5] {Numerically determined inverse participation ratio
    $P_{defect}(3T/4)$ (full line) versus scaled ac amplitude, for the same
    situation as considered in Fig.~3, compared to the high-frequency
    approximation~(\ref{ANA}) (dashed line).}  
\end{figure}

\begin{figure}
\caption[FIG.~6] {Average inverse participation ratios~(\ref{PAV}) for a
    driven, disordered lattice with $N = 101$ sites, as functions of the
    dimensionless ac amplitude $eF_1 d/\omega$, showing the remarkably sharp
    crossover from ``weakly'' to ``strongly'' localized states.
    The ac frequency is $\omega/\Delta = 1.0$, the dc field strength
    $e F_0 d/\omega = 1.0$. The disorder distribution is given by~(\ref{DIS}),
    with $\nu_{max}/\Delta = 0.01$, $0.05$, $0.10$, and $0.20$. The weaker the
    disorder, the sharper the spikes.}
\end{figure}

\begin{figure}
\caption[FIG.~7] {Quasienergy spectrum for a driven lattice with
    disorder strength $\nu_{max}/\Delta = 0.20$; all other parameters
    are as in Fig.~6.}
\end{figure}

\begin{figure}
\caption[FIG.~8] {Average inverse participation ratios for the same
    lattice as considered in Fig.~6, again for the ``one-photon resonance''
    $e F_0 d = 1 \cdot \omega$. The disorder strengths vary from
    $\nu_{max}/\Delta = 0.01$ (lowest curve) to $0.10$.}
\end{figure}

\begin{figure}
\caption[FIG.~9] {Same as Fig.~8, for $e F_0 d = 2 \cdot \omega$.}
\end{figure}

\begin{figure}
\caption[FIG.~10] {Same as Fig.~8, for $e F_0 d = 3 \cdot \omega$.}
\end{figure}

\begin{figure}
\caption[FIG.~11] {Boxes: characteristic ac field strengths $z_{1/2}$
    versus disorder strength for $n = 1$, $2$, and $3$, as determined from
    Figs.~8--10. Dashed lines: approximate relation~(\ref{APJ}) with
    $c_1 = 1.074$, $c_2 = 1.183$, and $c_3 = 1.189$. Full line: prediction
    of the full Anderson relation~(\ref{AND}) for $n=3$.}
\end{figure}

\begin{figure}
\caption[Fig.~12] {Time evolution of the center $\widehat{x}$ of a wave packet
    initially given by a Gaussian~(\ref{GAU}) in a disordered lattice.
    At $t = 0$ the packet is localized at the center of a lattice
    with $1001$ sites. The ac frequency is $\omega/\Delta = 1.0$,
    the dc field strength $eF_0d/\omega = 1.0$, and the ac field strength
    $eF_1d/\omega = 0.3$. The disorder strength is $\nu_{max}/\Delta = 0.1$.
    Without disorder, the packet would move to the left by 328 sites
    within $1000$~cycles $T = 2\pi/\omega$.}
\end{figure}

\begin{figure}
\caption[Fig.~13] {Time evolution of the width of the same wave packet
    as considered in Fig.~12 (full line). The initial width is
    $\sigma = 5 \, d$. The evolution of the width in the absence of
    disorder is shown by the dashed line; $\sigma$ would reach $33.2 \, d$
    after $1000$ cycles.}
\end{figure}

\begin{figure}
\caption[Fig.~14] {Average inverse participation ratio for a low-frequency
    field: $\omega/\Delta = 0.1$, with $eF_0d = 10 \, \omega$. The lattice
    has $101$ sites; the disorder strength is $\nu_{max}/\Delta = 0.1$.
    The sharp spike is at the first zero of $J_{10}$, where
    $eF_1d/\omega \approx 14.5$.}
\end{figure}

\begin{figure}
\caption[Fig.~15] {Quasienergy spectrum corresponding to the situation shown
    in Fig.~14.}
\end{figure}

\begin{figure}
\caption[Fig.~16] {Quasienergies for the two-level systems~(\ref{TBZ})
    with $D/\omega = 5.5$, and $eF_0d = \omega$. The interband matrix
    element is $X_{1,2}/d = -16/(9\pi^2)$, so that the resonance
    condition~(\ref{RCO}) is not satisfied. These are also the quasienergies
    for the two-band model~(\ref{TBA}) with $\Delta_1 = \Delta_2 = 0$.}
\end{figure}

\begin{figure}
\caption[Fig.~17] {Quasienergies for the two-band model~(\ref{TBA})
    with 20 sites. The parameters $D/\omega$, $eF_0d/\omega$, and $X_{1,2}/d$
    are as in Fig.~16; the bandwidths are $\Delta_1/\omega = 1.2$ and
    $\Delta_2/\omega = 1.0$. Arrows on top of the figure mark values of
    the scaled ac amplitude $eF_1d/\omega$ equal to zeros of the Bessel
    function $J_1$.}
\end{figure}

\begin{figure}
\caption[Fig.~18] {Same as Fig.~17, with reduced interband separation
    $D/\omega = 1.6$.}
\end{figure}

\begin{figure}
\caption[Fig.~19] {Quasienergies for the two-band model~(\ref{TBA})
    with $D/\omega = 3.5$, $\Delta_1/\omega = 1.2$, $\Delta_2/\omega = 1.0$,
    and $eF_0d = \omega$. Now the interband matrix element is
    $X_{1,2}/d = -0.5$, so that the resonance condition~(\ref{RCO}) is
    satisfied. Arrows on top again indicate zeros of $J_1$.
    The lattice consists of 20 sites.}
\end{figure}

\begin{figure}
\caption[Fig.~20] {Magnification of the first avoided quasienergy band
    crossing seen in Fig.~19.}
\end{figure}


\begin{references}
\bibitem{HarocheEtAl70} S. Haroche, C. Cohen-Tannoudji, C. Audoin,
    and J.P. Schermann,
    Phys. Rev. Lett. {\bf 24}, 861 (1970).
\bibitem{Chapman70} G.D. Chapman,
    J. Phys. B {\bf 3}, L36 (1970).
\bibitem{YabuzakiEtAl72} T. Yabuzaki, N. Tsukada, and T. Ogawa,
    J. Phys. Soc. Japan {\bf 32}, 1069 (1972).
\bibitem{YabuzakiEtAl74} T. Yabuzaki, S. Nakayama, Y. Murakami, and T. Ogawa,
    Phys. Rev. A {\bf 10}, 1955 (1974).
\bibitem{Shirley65} J.H. Shirley,
    Phys. Rev. {\bf 138}, B979 (1965).
\bibitem{Zeldovich67} Ya. B. Zel'dovich,
    Zh. Eksp. Theor. Fiz. {\bf 51}, 1492 (1966)
    [Sov. Phys. JETP {\bf 24}, 1006 (1967)].
\bibitem{Ritus67} V.I. Ritus,
    Zh. Eksp. Theor. Fiz. {\bf 51}, 1544 (1966)
    [Sov. Phys. JETP {\bf 24}, 1041 (1967)].
\bibitem{Sambe73} H. Sambe,
    Phys. Rev. A {\bf 7}, 2203 (1973).	    
\bibitem{CohenTannoudji68} C. Cohen-Tannoudji,
    in {\em Carg\`ese Lectures in Physics}, ed. M. L\'evy, Vol.\ 2, p.\ 347
    (Gordon and Breach, New York, 1968).
\bibitem{CohenTannoudji92} C. Cohen-Tannoudji, J. Dupont-Roc, and G. Grynberg,
    {\em Atom-Photon Interactions}
    (John Wiley \& Sons, New York, 1992). 
\bibitem{PeggSeries70} D.T. Pegg and G.W. Series,
    J. Phys. B {\bf 3}, L33 (1970).
\bibitem{FEL1} P.S.S. Guimar\~aes, B.J. Keay, J.P. Kaminski, S.J. Allen Jr.,
    P.F. Hopkins, A.C. Gossard, L.T. Florez, and J.P. Harbison,
    Phys. Rev. Lett. {\bf 70}, 3792 (1993).
\bibitem{IgnatovEtAl94} A.A. Ignatov, E. Schomburg, K.F. Renk, W. Schatz,
    J.F. Palmier, and F. Mollot,
    Ann. Physik {\bf 3}, 137 (1994).	
\bibitem{FEL2} B.J. Keay, S.J. Allen Jr., J. Gal\'an, J.P. Kaminski, 
    K.L. Campman, A.C. Gossard, U. Bhattacharya, and M.J.W. Rodwell,
    Phys. Rev. Lett. {\bf 75}, 4098 (1995).	   
\bibitem{FEL3} B.J. Keay, S. Zeuner, S.J. Allen Jr., K.D. Maranowski,
    A.C. Gossard, U. Bhattacharya, and M.J.W. Rodwell,
    Phys. Rev. Lett. {\bf 75}, 4102 (1995). 
\bibitem{DakhnovskiiBavli93} Y.~Dakhnovskii and R.~Bavli,
    Phys.~Rev.~B {\bf{48}}, 11010 (1993).
\bibitem{GorbatsevichEtAl95} A.A. Gorbatsevich, V.V. Kapaev, and Yu. V. Kopaev,
    Zh. Eksp. Teor. Fiz. {\bf 107}, 1320 (1995)
    [Sov. Phys. JETP {\bf 80}, 734 (1995)].
\bibitem{GrossmannEtAl91a} F. Grossmann, P. Jung, T. Dittrich, and P. H\"anggi,
    Z. Phys. B {\bf 84}, 315 (1991).
\bibitem{GrossmannEtAl91b} F. Grossmann, T. Dittrich, P. Jung, and P. H\"anggi,
    Phys. Rev. Lett. {\bf 67}, 516 (1991).
\bibitem{Haenggi95} For a recent review, see
    P. H\"anggi,
    in {\em Quantum Dynamics of Submicron Structures},
    ed. H.A. Cerdeira et al.\
    (Kluwer, Dordrecht, 1995).
\bibitem{LlorentePlata92} J.M. Gomez Llorente and J. Plata,
    Phys. Rev. A {\bf 45}, R6958 (1992).
\bibitem{GrossmannHaenggi92} F. Grossmann and P. H\"anggi,
    Europhys. Lett. {\bf 18}, 571 (1992).
\bibitem{Holthaus92a} M. Holthaus,
    Phys. Rev. Lett. {\bf 69}, 1596 (1992).     
\bibitem{Howland92} J.S. Howland,
    J. Phys. A {\bf 25}, 5177 (1992).
\bibitem{FukuyamaEtAl73} H. Fukuyama, R.A. Bari, and H.C. Fogedby,
    Phys. Rev. B {\bf 8}, 5579 (1973).
\bibitem{Bloch28} F. Bloch,
    Z. Phys. {\bf 52}, 555 (1928).
\bibitem{Zener34} C. Zener,
    Proc. R. Soc. London A {\bf 145}, 523 (1934).        
\bibitem{Houston40} W.V. Houston,
    Phys. Rev. {\bf 57}, 184 (1940).
\bibitem{Chu89} For a review, see
    S.-I Chu,
    Adv. Chem. Phys. {\bf 73}, 739 (1989).
\bibitem{HolthausHone93} M. Holthaus and D.W. Hone,
    Phys. Rev. B {\bf 47}, 6499 (1993).
\bibitem{Zak93} J. Zak,
    Phys. Rev. Lett. {\bf 71}, 2623 (1993).
\bibitem{DunlapKenkre86} D.H. Dunlap and V.M. Kenkre,
    Phys. Rev. B {\bf 34}, 3625 (1986).
\bibitem{Zhao91} X.-G. Zhao,
    Phys. Lett. A {\bf 155}, 299 (1991).
\bibitem{ShonNazareno92} N. Hong Shon and H.N. Nazareno,
    J. Phys.: Condens. Matter {\bf 4}, L611 (1992).
\bibitem{IgnatovEtAl95} A.A. Ignatov, E. Schomburg, J. Grenzer, K.F. Renk,
    and E.P. Dodin,
    Z. Phys. B {\bf 98}, 187 (1995).
\bibitem{MeierEtAl95b} T. Meier, F. Rossi, P. Thomas, and S.W. Koch, 
    Phys. Rev. Lett. {\bf 75}, 2558 (1995).	
\bibitem{Holthaus92b} M. Holthaus,
    Phys. Rev. Lett. {\bf 69}, 351 (1992).
\bibitem{Holthaus92c} M. Holthaus,    
    Z. Phys. B {\bf 89}, 251 (1992).
\bibitem{HoneHolthaus93} D.W. Hone and M. Holthaus,
    Phys. Rev. B {\bf 48}, 15123 (1993).
\bibitem{Anderson58} P.W. Anderson,
    Phys. Rev. {\bf 109}, 1492 (1958).
\bibitem{MottTwose61} N.F. Mott and W.D. Twose,
    Adv. Phys. {\bf 10}, 107 (1961).
\bibitem{Thouless74} D.J. Thouless,
    Phys. Rep. {\bf 13}, 93 (1974).
\bibitem{KramerMacKinnon93} B. Kramer and A. MacKinnon,
    Rep. Prog. Phys. {\bf 56}, 1469 (1993).
\bibitem{Thouless79} D.J. Thouless,
    in: {\em Ill-Condensed Matter: Les Houches Session XXXI},
    ed. R. Balian, R. Maynard, and G. Toulouse,
    (North Holland, Amsterdam, 1979).
\bibitem{HolthausEtAl95a} M. Holthaus, G.H. Ristow, and D.W. Hone,
    Phys. Rev. Lett. {\bf 75}, 3914 (1995).    
\bibitem{HolthausEtAl95b} M. Holthaus, G.H. Ristow, and D.W. Hone,
    Europhys. Lett. {\bf 32}, 241 (1995).
\bibitem{RotvigEtAl95} J. Rotvig, A.-P. Jauho, and H. Smith,
    Phys. Rev. Lett. {\bf 74}, 1831 (1995).
\bibitem{HoneZhao95} D.W. Hone and X.-G. Zhao,
    {\em Time Periodic Behavior of Multiband Superlattices in Static Electric
    Fields}, preprint (1995). 
\bibitem{HolthausHone94} M. Holthaus and D.W. Hone,
    Phys. Rev. B {\bf 49}, 16605 (1994).
\bibitem{DreseHolthaus95} K. Drese and M. Holthaus,
    J. Phys.: Condens. Matter, in press (1996).
\bibitem{ZhaoEtAl95} X.-G. Zhao, R. Jahnke, and Q. Niu,
    Phys. Lett. A {\bf 202}, 297 (1995).
\bibitem{Buettiker93} M. B\"{u}ttiker,
    J. Phys.: Condens. Matter {\bf 5}, 9361 (1993).
\bibitem{Buettiker94} M. B\"{u}ttiker,
    {\em Time-Dependent Current Partition in Mesoscopic Conductors},
    Preprint UGVA-DPT 1994/10-863 (1994).    
\bibitem{MeierEtAl94} T. Meier, G. von Plessen, P. Thomas, and S.W. Koch,
    Phys. Rev. Lett. {\bf 73}, 902 (1994).
\bibitem{MeierEtAl95a} T. Meier, G. von Plessen, P. Thomas, and S.W. Koch,
    Phys. Rev. B {\bf 51}, 14490 (1995).
\bibitem{MooreEtAl94} F.L. Moore, J.C. Robinson, C. Bharucha, P.E. Williams,
    and M.G. Raizen,
    Phys. Rev. Lett. {\bf 73}, 2974 (1994).
\bibitem{BardroffEtAl95} P.J. Bardroff, I. Bialynicki-Birula, D.S. Kr\"ahmer,
    G. Kurizki, E. Mayr, P. Stifter, and W.P. Schleich,
    Phys. Rev. Lett. {\bf 74}, 3959 (1995).
\bibitem{RobinsonEtAl95} J.C. Robinson, C. Bharucha, F.L. Moore, R. Jahnke,
    G.A. Georgakis, Q. Niu, M.G. Raizen, and B. Sundaram,
    Phys. Rev. Lett. {\bf 74}, 3963 (1995).
\bibitem{LatkaWest95} M. Latka and B.J. West,
    Phys. Rev. Lett. {\bf 75}, 4202 (1995).
\bibitem{BenDahanEtAl96} M. Ben Dahan, E. Peik, J. Reichel, Y. Castin,
    and C. Salomon,
    {\em Bloch Oscillations of Atoms in an Optical Potential},
    preprint, submitted to Phys. Rev. Lett. (1996).
\end{references}
\end{document}